\documentclass[journal]{IEEEtran}

\usepackage{times}
\usepackage{xcolor}
\usepackage{epsfig}
\usepackage{graphicx}
\usepackage{amssymb}
\usepackage{multicol,multirow}
\usepackage{booktabs}
\usepackage{bigstrut}
\usepackage{floatrow}
\usepackage{rotating}
\usepackage{adjustbox}
\usepackage{amsmath}

\ifCLASSINFOpdf

\else

\fi

\begin{document}
\title{ESDMR-Net: A Lightweight Network With Expand-Squeeze and Dual Multiscale Residual Connections for Medical Image Segmentation}
\author{Tariq M. Khan,~\IEEEmembership{Member,~IEEE,}
        Syed S. Naqvi,~\IEEEmembership{Member,~IEEE,}
        and~Erik Meijering,~\IEEEmembership{Fellow,~IEEE}
\thanks{Tariq M. Khan and  Erik Meijering are with the School of Computer Science and Engineering, University of New South Wales, Sydney, NSW, Australia}
\thanks{Syed S. Naqvi is with the Department of Electrical and Computer Engineering, COMSATS University Islamabad, Pakistan.}}

\maketitle

\begin{abstract}
Segmentation is an important task in a wide range of computer vision applications, including medical image analysis. Recent years have seen an increase in the complexity of medical image segmentation approaches based on sophisticated convolutional neural network architectures. This progress has led to incremental enhancements in performance on widely recognised benchmark datasets. However, most of the existing approaches are computationally demanding, which limits their practical applicability. This paper presents an expand-squeeze dual multiscale residual network (ESDMR-Net), which is a fully convolutional network that is particularly well-suited for resource-constrained computing hardware such as mobile devices. ESDMR-Net focusses on extracting multiscale features, enabling the learning of contextual dependencies among semantically distinct features. The ESDMR-Net architecture allows dual-stream information flow within encoder-decoder pairs. The expansion operation (depthwise separable convolution) makes all of the rich features with multiscale information available to the squeeze operation (bottleneck layer), which then extracts the necessary information for the segmentation task. The Expand-Squeeze (ES) block helps the network pay more attention to under-represented classes, which contributes to improved segmentation accuracy. To enhance the flow of information across multiple resolutions or scales, we integrated dual multiscale residual (DMR) blocks into the skip connection. This integration enables the decoder to access features from various levels of abstraction, ultimately resulting in more comprehensive feature representations. \color{black}We present experiments on seven datasets from five distinct examples of applications: segmentation of retinal vessels ($2\times)$, skin lesions ($2\times)$, digestive tract polyps, lung regions, and cells. Our model demonstrates strong performance, with an F1 score of 0.8287\%, 0.8211\%, 0.9034\%, 0.9451\%, 0.9543\%, 0.9840\%, and 0.8424\% on the DRIVE, CHASE, ISIC2017, ISIC2016, CVC-ClinicDB, MC and MoNuSeg datasets, respectively. Remarkably, our model achieves these results despite having significantly fewer trainable parameters, with a reduction of two or even three orders of magnitude. \color{black}
\end{abstract}


\section{Introduction}

The accurate segmentation of anatomical structures and abnormalities in medical images plays a vital role in the diagnosis and, ultimately, in the provision of appropriate treatment for patients. The segmentation process can be complex, even for highly experienced human experts in the field \cite{naveed2021towards,imtiaz2021screening, soomro2018impact, khan2017efficient}. Challenges arise due to factors such as ambiguous structural boundaries, inherent uncertainty in segmented regions, diverse textures, nonuniform intensity distribution, and significant contrast variations commonly encountered in medical images \cite{khan2021residual,khan2022width,khan2021rc,aslam2022ensemble,arsalan2022prompt,iqbal2023mlr}. These complexities emphasise the importance of developing advanced segmentation techniques to facilitate clinical diagnosis.

Current techniques for medical image segmentation can be broadly categorised into unsupervised and supervised approaches. Unsupervised approaches often rely on low-level features and ad-hoc rules, which can lead to limited generalisation capabilities. In contrast, supervised methods use human-annotated training images, which typically leads to superior segmentation accuracy compared to unsupervised techniques. Deep learning has emerged as the most effective among supervised approaches, facilitating end-to-end segmentation with enhanced accuracy and generalisation. Several convolutional neural networks (CNNs) have been specifically designed and developed for medical image segmentation, demonstrating the versatility and potential of these techniques within the field of medical imaging \cite{khan2021residual,khan2022leveraging,khan2022t,iqbal2022g,khan2022mkis,khan2023simple,naqvi2023glan,khan2023retinal,khan2023feature,abbasi2023lmbis,iqbal2023ldmres}.

To maintain spatial localisation information, fully convolutional networks (FCNs) with skip layers were suggested for semantic segmentation \cite{Jonathan2015}. U-Net, which was inspired by FCNs, has become a prominent network architecture for a wide range of segmentation applications \cite{Ronneberger2015}. CNN models that learn new information through a reinforcement strategy technique have also been proposed \cite{guo2018retinal}. The introduction of a connection-sensitive U-Net has further improved the segmentation results \cite{li2019connection}. Using a combination of higher resolution features and upsampled features, it makes use of contextual information from deeper layers. A joint loss at the segmentation level and at the pixel level has been proposed to address this problem \cite{yan2018joint}. An evaluation metric for skeleton similarity has been developed to evaluate the skeleton map generated by combining reference and source segmentation maps \cite{yan2017skeletal}.


In situations where computing resources are limited, it becomes crucial to minimise the number of trainable parameters. For example, in point-of-care (POC) diagnostics, which aims to perform medical tests at or near the point in time and space where the patient receives care, there may only be room for lightweight devices, such as smartphones \cite{Xu2015POC}. Smartphone-based POC diagnostics and decision support systems have been explored for the analysis of the retinal vasculature and the diagnosis of retinal disease \cite{xu2016smartphone,BOUROUIS2014retina}. However, current state-of-the-art methods for medical image segmentation, including encoder-decoder-based networks and transformer-based methods, are not practical in POC diagnostics, as they often require excessive computing resources.

\textcolor{black}{One approach to reducing the number of trainable parameters is to reduce the number of filters used in each layer of the network} and/or the number of hidden layers. Consequently, shallow networks often emerge as a feasible alternative to their deep-network counterparts \cite{9207668}. These networks can provide a balance between computational efficiency and performance, making them suitable for applications on devices with limited resources. However, the feature maps they produce may contain overlapping information if the variation in the features is low compared to the number of filters employed \cite{app9010108}.

In the literature, a less complex alternative is to use lightweight networks \cite{Howard_2019_ICCV,khan2022t}. To comply with hardware limitations, these networks often reduce the number of filters per layer \cite{Howard_2019_ICCV}. For example, MobileNet-V3 \cite{Howard_2019_ICCV} has a large number of pooling kernels and a large stride, which limit its memory footprint but contribute to its low performance when segmenting high-frequency features. T-Net \cite{khan2022t}, GLAN \cite{naqvi2023glan}, G-Net \cite{iqbal2022g}, PLVS-Net \cite{arsalan2022prompt}, and MKIS-Net \cite{khan2022mkis}, on the other hand, employ small filter kernels to extract low-frequency information, but due to their shallow structures, they perform poorly when dealing with datasets that exhibit significant feature variation. \textcolor{black}{Hence, an architecture is required that can extract robust features for all inherent frequencies in medical images, encode feature variations for varying datasets and conform with the computational requirements of POC diagnosis.} 

This paper presents a novel, efficient and lightweight design that we refer to as ESDMR-Net, short for the expand-squeeze dual multiscale residual network. \textcolor{black}{ESDMR-Net surpasses MobileNet-V3 in successfully capturing the essential high-frequency features necessary for medical image segmentation, thus overcoming the shortcomings of MobileNet-V3 in this aspect. ESDMR-Net also outperforms T-Net by solving the latter's shortcomings in managing large feature variation, providing a more flexible solution for a wider range of datasets. ESDMR-Net is a generic architecture designed specifically for environments with limited memory and computing resources.}

\textcolor{black}{The proposed network design involves expand-squeeze (ES) blocks that are used on multiple scales to capture high-frequency information and cross-channel correlations between features to exploit class information and promote under-represented classes, while separable convolution in depth is used in the ES blocks to accumulate more useful features with low computational requirements.
To ensure that rich feature representations are preserved in the decoder, a dual multiscale residual (DMR) block is proposed at the skip connections. DMR blocks in skip connections facilitate effective transmission of gradient information on multiple scales so that feature information is preserved at the decoder end. The DMR expands the receptive field of the decoder, allowing it to access more contextual information from different levels of the encoder. The DMR blocks aim to address challenges such as loss of feature information in the transmission process and the vanishing gradient problem, and they improve the network's ability to capture and represent information from different levels of abstraction. To deal with highly varying features, the network is designed to have a deep structure by utilising a large number of branches and the reuse of ES blocks so that the refinement of the features can be accumulated through the network.}

The proposed ESDMR-Net design has only around 0.7 million parameters, which is the same as M2U-Net but five times less than MobileNet-V3. ESDMR-Net overcomes the limitations of existing lightweight networks such as MobileNet-V3 \cite{Howard_2019_ICCV} and T-Net \cite{khan2022t}. It is a generic architecture designed specifically for environments with limited memory and computing resources. 


The main contributions are as follows:

\begin{itemize}
\item A novel deep convolutional network architecture with ES blocks is proposed for improved feature representation \textcolor{black}{to cater high frequency and feature variations in medical images.}
\item The use of skip connections and DMR blocks \textcolor{black}{enable feature preservation at the decoder end, gradient propagation, preservation of contextual information, and allows fine-grained feature reuse.}
\item The deep structure of the network allows it to \textcolor{black}{be robust to highly varying features via feature refinement and accumulation}.
\item \textcolor{black}{The suggested framework is tailored for limited resource settings, boasting a low resource footprint, yet it maintains competitive performance compared to existing methodologies.}
\item The proposed network has been evaluated on five different applications and seven distinct datasets to show its robustness and generalizability.
\end{itemize}

\textcolor{black}{The remainder of the paper is organised as follows: Section \ref{sec:Related Work} discusses recent relevant works. The design and technical details of the proposed ESDMR-Net are provided in Section \ref{networks}. The experimental setup is summarised in Section \ref{experimentalResults}. The detailed experimental results of ESDMR-Net and comparisons with other methods are presented in Section \ref{Results}. Finally, Section \ref{Conclusions} discusses the main findings and conclusions of the proposed work.}

\section{Related Work}
\label{sec:Related Work}
Before presenting our proposed method, we first provide a concise overview of state-of-the-art methods that serve as benchmarks for evaluating our method. Our evaluation spans diverse applications, ensuring comprehensive comparisons based on the performances reported by both original developers and existing literature.

Among the lightweight architectures, \cite{OLIVEIRA2018229} presented a fully convolutional low parameter network (FCN) that, while efficient, can experience information loss during upsampling. In \cite{khan2022t} a reduced complexity architecture, T-Net was proposed that introduced a trade-off between representational learning and computational overhead. MobileNet-V3, a platform-aware network that uses the NetAdapt algorithm, was introduced in \cite{Howard_2019_ICCV}. In \cite{iqbal2022g}, G-Net, a lightweight variant of GoogleNet, was suggested that emphasises the preservation of spatial context and the reduction of complexity. An encoder-decoder based lightweight architecture named GLAN was proposed in \cite{naqvi2023glan}. MKIS-Net \cite{khan2022mkis}, a lightweight network designed for image segmentation, employed multiple kernels to create an effective receptive field, thus enhancing the performance of segmentation. \color{black} These architectures optimise for reduced complexity, balancing the trade-off between model size and performance. However, segmentation quality can be compromised if information is lost during upsampling in lightweight architectures such as the lightweight FCN \cite{OLIVEIRA2018229}. The effectiveness of lightweight architectures may depend on the task and may not outperform more complex models in all scenarios. \color{black}

The encoder-decoder category showcases notable designs such as SegNet \cite{khan2020residual}, which employs an encoder-decoder structure with skip connections for pixel-wise classification. Classic U-Net \cite{10.1007/978-3-319-24574-4_28} effectively captures the global and local context through skip connections, maintaining fine details. U-Net++, an extension of UNet, improves segmentation accuracy by nesting densely connected pathways. MobileNetV2 components were leveraged in the encoder-decoder architecture of M2U-Net by \cite{LaibacherWJ19_CVPRW}. Attention gates were incorporated into Attention Res-UNet to selectively emphasise relevant features \cite{maji2022attention}. MMS-Net \cite{khan2023simple} is specifically designed for pixel-level detection, incorporating multipath and multiscale convolutional operations, as well as multiple deep feature aggregation techniques. \textcolor{black}{PLVS-Net was proposed for accurate segmentation of retinal vessels by \cite{arsalan2022prompt}. Using prompt blocks, PLVS-Net achieves improved feature extraction and segmentation performance with fewer than a million parameters, making it suitable for resource-constrained devices. The prompt blocks extract and preserve useful features and relay them to the decoder to bridge the semantic gap between the encoder and the decoder.} RCED-Net, which addresses the segmentation issue by approaching it as a pixel-wise semantic segmentation task, was proposed by \cite{khan2020residual}. \color{black}
A modified version of the U-Net framework for performing segmentation on chest radiographs was proposed by \cite{Kumarasinghe_Kolonne_Fernando_Meedeniya_2022}. Subsequently, segmented images were used for classification purposes to assess the resulting influence on performance. The segmentation model proposed by \cite{Shyamalee2022} is based on an attention U-Net with three separate convolutional neural networks (CNNs) backbones: visual geometry group 19 (VGG19), Inception-v3, and residual neural network 50 (ResNet50). 
A U-Net model with spatial attention (SA-UNet) architecture was proposed for the segmentation of chest X-ray (CXR) images by \cite{10050951}. Attention U-Net models with three architectures of convolutional neural networks (CNN), namely Inception-v3, Visual Geometry Group 19 (VGG19) and Residual Neural Network 50 (ResNet50), were proposed for segmentation of fundus images by \cite{9765303}. 
Edward's probabilistic programming language, together with the U-Net model and adversarial generative networks employing different optimisers, was proposed by \cite{Rubasinghe_Meedeniya_2019} for ultrasound nerve segmentation. These encoder-decoder designs can capture global and local context effectively, providing good pixel-wise classification. Handling the semantic gap between encoders and decoders can be difficult, limiting the model's capacity to capture complex features. Most of these systems are highly dependent on skip connections, which can cause difficulty in obtaining detailed features in some cases.

\color{black}
The integration of contextual information was explored by \cite{Wang2019} with DEU-Net, which combines dual encoders with spatial and context paths, employing feature fusion and channel attention for improved accuracy. Vessel-Net \cite{Wu2019}, used multiscale multipath supervision with initial residual blocks for enhanced segmentation. SFA \cite{Fang2019} is characterised by selective aggregation of features through common and paired decoders. An MS-NFN network, proposed by \cite{Wu2018}, incorporates multiscale information at the patch level to enhance performance. In \cite{Arsalan2019}, a semantic segmentation network is proposed that uses a dual-residual-stream-based architecture. The first stream is a sequential path that connects adjacent encoder and decoder layers with non-identity mappings, while the second stream is an external path that directly connects encoder and decoder layers to capture edge information (high-frequency spatial features). \color{black} {One advantage of these methods that use contextual information is improved segmentation accuracy. However, as observed in DEU-Net, the additional complexity of contextual information integration could influence real-time processing and require more computational resources. The nature of the segmentation challenge can influence the effectiveness of tasks that involve feature aggregation and multiscale information.} 

\color{black}
The pyramidal architecture, proposed by \cite{9049412}, was employed for the effective integration of global and multiscale context information. 
DAGAN-Net, proposed by \cite{FANG2023109248}, comprises two main modules: the UNet-SCDC segmentation module and the DD module. 
FAT-Net \cite{WU2022102327} enhances overall performance by activating key channels and suppressing irrelevant background noise. For effective and precise segmentation of skin lesions, \cite{DAI2022102293} proposed a novel network architecture known as Ms RED.
A Swin-UNet \cite{cao2023swin} uses a transformer-based architecture for medical image segmentation. A novel segmentation scheme known as UCTransNet \cite{UCTransNet2022}, which incorporates a unique CTrans module into U-Net, focussing on the channel perspective and employing an attention mechanism. \color{black} Pyramidal designs can effectively combine global and multiscale context information, resulting in enhanced segmentation. Pyramidal structures may have a high computational cost, which limits their use in resource-constrained contexts. Specialised techniques may excel at specific tasks, but may lack adaptability when applied to a wider range of segmentation challenges. \color {black}

DRIU \cite{10.1007/978-3-319-46723-8_17} is designed specifically for the segmentation of vascular networks and optical discs, incorporating specialised layers.
PraNet \cite{Fan2020} used a parallel reverse attention network for precise polyp segmentation, to improve the precision of segmentation. 
AS-Net \cite{HU2022117112} combines spatial and channel attention mechanisms to enhance the performance of segmentation for skin lesions. The innovative BiO-Net \cite{BiO-Net2020} used a bidirectional O-shape network that cyclically employs its building blocks, offering versatile domain enhancement. 
ERFNet, proposed by \cite{Romera_2018}, was designed for the semantic segmentation of real-world images with a large number of classes. The network employs a significant number of convolutional filters to represent the features at each stage.
AACA-MLA-D-Net, proposed by \cite{yuan2021multi}, efficiently leverages low-level details and complementary information from multiple layers to accurately differentiate vessels from the background while maintaining minimal model complexity.

These methodological variations provide a comprehensive foundation for evaluating our proposed method's performance across diverse domains.

\section{Proposed Network Design}\label{networks}

\begin{figure*}[ht]
  \centering
  \includegraphics[width=1\textwidth]{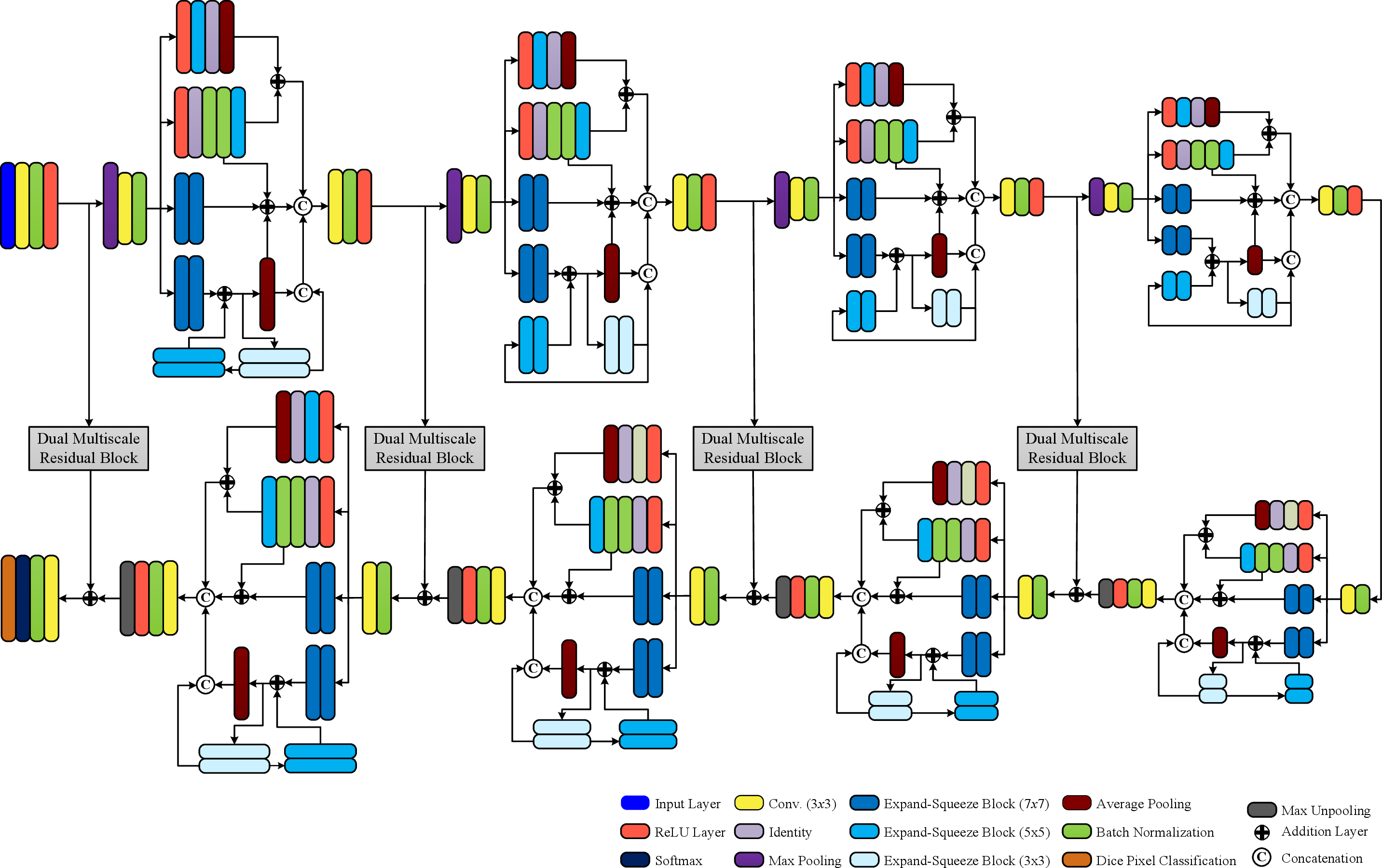}
  \caption{Block diagram of the proposed ESDMR-Net. The input is first processed by an input block consisting of convolution (Conv), batch normalisation, and rectified linear unit (ReLU) activation. Subsequently, four encoder and four decoder blocks further extract multiscale information. The final processing by an output block consisting of convolution, batch normalisation, softmax activation, and Dice pixel classification produces the output segmentation map. DMR blocks in the skip connections between the encoder and decoder blocks, as well as the input and output blocks, facilitate the preservation/restoration of information and training of the network.}
  \label{fig:ESED}
\end{figure*}

The proposed ESDMR-Net encoder-decoder architecture (Fig.~\ref{fig:ESED}) has been specifically designed to extract, enhance, and retain comprehensive feature details at each phase of the encoding-decoding process, while taking into account computational budget constraints, without compromising performance. \color{black}The extracted features can be broadly categorised into low-level (multiscale frequency), mid-level (class-specific), and high-level (broader contextual information) features\color{black}. Using a compound scaling approach \cite{tan2019efficientnet}, we constructed our network to be deep, with more than 100 layers. The ESDMR-Net comprises four encoder and four matching decoder blocks. It captures multiscale information through receptive fields of different scales. Average pooling is used to downsample the feature maps in each encoder block, whereas bilinear upsampling is used at the decoder end. Throughout the network, a pooling kernel with dimensions of 3$\times$3 and a stride of 1 is utilised.

\textcolor{black}{The internal architecture of an ESDMR-Net encoder block has multiple branches featuring separable depth convolution with varying receptive fields to capture a rich collection of features of varying frequency. The rationale for employing depthwise separable convolution is based on empirical evidence that disentangling cross-channel correlations from spatial correlations can lead to the extraction of more meaningful features \cite{Guo2019Depthwise}. Moreover, the ability of depthwise separable convolution to capture more concepts compared to its counterparts is also supported by previous work \cite{Guo2019Depthwise}.} The network decoder blocks follow the same topology as the encoder block, but with the downsampling operation replaced by upsampling. A highlight of the architecture is the ES block, which, unlike the Fire module \cite{iandola2016squeezenet} features an expansion followed by a squeeze operation. The rationale behind this strategy is that \textcolor{black}{the expansion operation extracts important features containing multiscale information, which can be subsequently used by the squeeze or bottleneck layer to extract compact salient information.}. Furthermore, overparameterization has been shown to increase the efficiency of compact networks \cite{Guo2020ExpandNets}, and expanded features improve the expressiveness of the network \cite{Yu2020bmvc}. Prior work has explored the ES convolution for superresolution \cite{Zhang2021SR,Wang2022SR}, which generally features a $3\times 3$ convolution layer for expansion followed by a $1 \times 1$ squeeze layer (bottleneck). In contrast, the proposed ES block employs expansion layers of varying filter sizes to account for features occurring at varying scales. Specifically, a depthwise separable convolution is employed for increased expressiveness through expansion while simultaneously ensuring compactness. In addition, the ES block incorporates nonlinear and normalisation operations for nonlinearity training and faster convergence. The proposed scale-aware approach is particularly well-suited to medical image segmentation, where the regions of interest occur at varying scales and have varied boundary structures.


\begin{figure}[!t]
  \centering
  \includegraphics[width=0.25\textwidth]{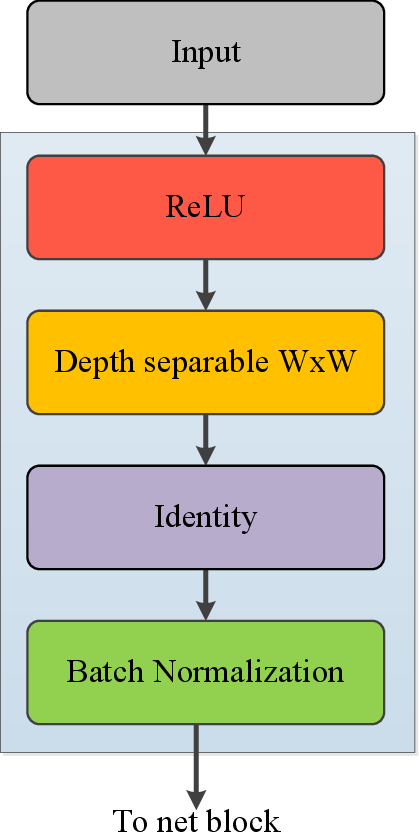} 
  \caption{Internal architecture of the ES block. The input is processed successively by rectified linear unit (ReLU) activation, an expansion layer implemented by depthwise separable convolution of specified scale $W \times W$ pixels, a squeeze layer implemented by identity mapping using $1\times1$ convolution, and batch normalization.}
  \label{fig:ESB}
\end{figure}

\begin{figure}
  \centering
  \includegraphics[width=0.9\textwidth]{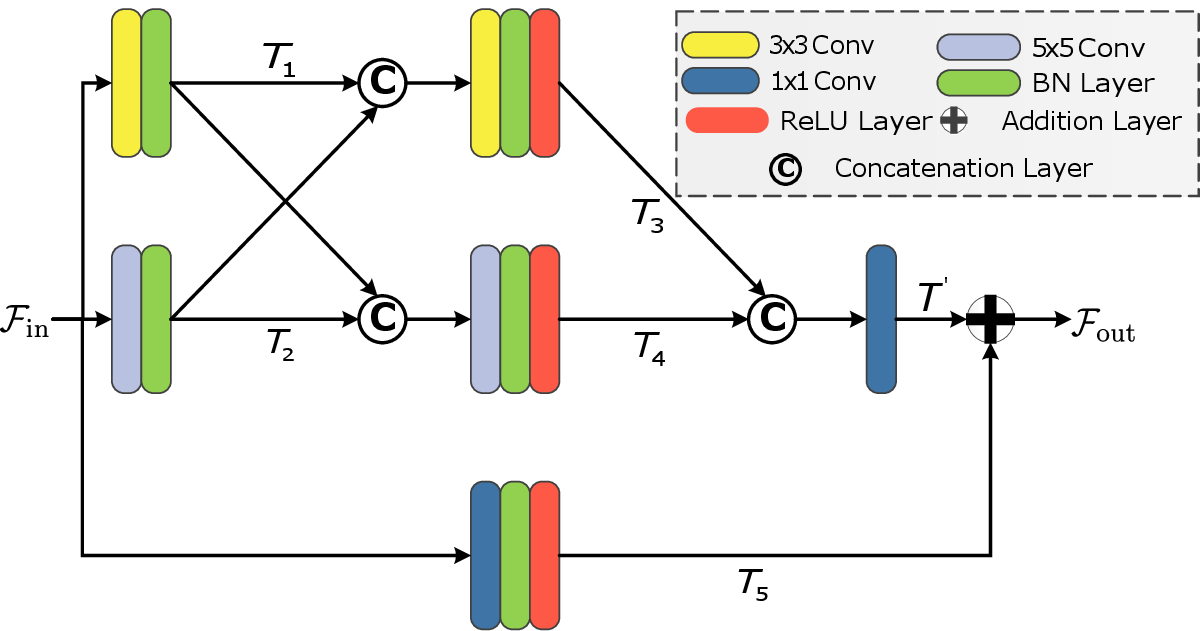}
  \caption{\color{black}Schematic diagram of the dual multiscale residual (DMR) block.\color{black}}
  \label{DMRBlock}
\end{figure}

\subsection{Expand-Squeeze Block}

The ES block consists of an expansion layer followed by a squeeze or the representational bottleneck layer (Fig.~\ref{fig:ESB}). The expansion operation is responsible for capturing spatial correlations and cross-channel correlations between features. The squeeze layer then extracts the salient information in a compact feature representation. 
\textcolor{black}{By capturing the interdependencies among channels, the proposed ES block enables the network to acquire more distinctive features for each class and dynamically adjust the channel-wise feature responses. The expansion operation enables the proposed ES block to capture class-specific features, and the squeeze layer extracts salient information to promote under-represented classes. This helps the proposed ES module combat class imbalance issues.} 
Therefore, the ES block used at different spatial scales allows the network to balance both the feature capacity and the expressiveness at the same time.

The proposed approach is different from the Fire module \cite{iandola2016squeezenet}, as feature correlations and multiscale information are learnt before transforming the output to a new channel space. In contrast to the Inception module \cite{Szegedy2016}, the expansion layer is implemented with depthwise separable convolution to extract multiscale features with high representational power and low computational overhead. By capturing the interdependencies among channels, the proposed ES block enables the network to acquire more distinctive features and dynamically adjust channel-wise feature responses. This helps to improve image segmentation performance.

In this work, the expansion operation is performed by depthwise separable convolution, where kernel sizes of $3\times3$, $5\times5$, and $7\times7$ pixels are employed to capture spatial correlations at multiple scales. Let ${\widehat{K}} \in \mathbb{R}^ {H \times W \times M}$ be a depthwise convolution kernel, where $H$ and $W$ are the spatial dimensions of the kernel and $M$ is the number of channels. The output ${D} \in \mathbb{R}^ {K \times L \times M}$ of the intermediate depthwise convolution given a feature map ${F} \in \mathbb{R}^ {K \times L \times M}$ with spatial dimensions $K \times L$ pixels can be computed as:
\begin{equation}
{D}_{k,l,m} = \sum_{p,q} {\widehat{K}}_{p,q,m} {F}_{k-p,l-q,m},
\end{equation}
where a convolution filter per input channel is applied to obtain an $M$-channel output. Given this output and multiple pointwise kernels, ${\dot{K}}_n \in \mathbb{R}^ {1 \times 1 \times M}$, the output of the pointwise convolution, which essentially computes a weighted linear combination of the input channels, is defined as:
\begin{equation}
{E}_{k,l,n} = \sum_{m} {\dot{K}}_{m,n} {D}_{k,l,m}.
\end{equation}
To extract and consolidate spatial and channel information from the expansion layers, a $1 \times 1$ convolution is employed in our ES block, which is similar to a bottleneck or squeeze operation. Both the expand and squeeze operations enable the encoder block to encapsulate useful features with low computational overhead.


\subsection{Dual Multiscale Residual Block}

\textcolor{black}{The DMR block (Fig.~\ref{DMRBlock}) is designed to extract and preserve features that encompass a range of spatial scales, from fine-grained details to broader contextual information. When used in conjunction with skip connections, it serves as an effective method for feature extraction and refinement by leveraging multiscale information.} Inclusion of DMR blocks in skip connections facilitates effective transmission of gradient information on multiple scales. This assists in mitigating the vanishing gradient issue, consequently enhancing the network's convergence and stability. It also expands the receptive field of the decoder, allowing it to access more contextual information from different levels of the encoder. \textcolor{black}{This helps in capturing larger structures and contextually rich representations. DMR blocks facilitate the reuse of features on different scales, allowing the network to take advantage of local and global information for better feature representation. The robustness of the features extracted by the proposed encoder and the preservation of the features by using DMR blocks in skip connections are supported by the results of our ablation study (Section \ref{Results}).}

In this setup, the encoder's feature maps before the max-pooling are fed as input to the DMR block, whereas the skip connections enable the preservation of essential information between layers. Given three layers of operation (Fig.~\ref{DMRBlock}), the internal processing of a DMR block can be defined as:
\begin{align} 
T_1 &= \beta_n\left ( f^{3\times3}_1 \ast \mathcal{F}_\text{in} \right ), \\
T_2 &= \beta_n\left ( f^{5\times5}_1 \ast \mathcal{F}_\text{in} \right ), \\
T_3 &= \Re\left ( \beta_n\left ( f^{3\times3}_2 \ast \left\{ T_1,T_2 \right\} \right ) \right ),  \\
T_4 &= \Re\left ( \beta_n\left ( f^{5\times5}_2 \ast \left\{ T_2,T_1 \right\} \right ) \right ),  \\ 
T^{'} &=  f^{1\times1}_3 \ast \left\{ {T4,T3}\right\},  
\label{Eqn:dmrop}
\end{align}
where $\mathcal{F}_\text{in}$ is the input feature tensor, $f$ represents the convolution kernels with kernel size depicted as superscripts and layer number as subscript, $\ast$ represents the convolution operation, $\beta_n$ denotes the batch normalization operation and the ReLU operation is depicted by $\Re(x) = \max(0,x)$. Finally, $\left\{\cdot,\cdot\right\}$ denotes the concatenation of features.





The unique dual-cross feature sharing of the proposed DMR block ensures robust fusion and detection at multiple scales. The output of the DMR block is computed as:
\begin{equation}
\mathcal{F}_\text{out} = T^{'} \bigoplus \; \Re\left ( \beta_n\left ( f^{1\times1}_2 \ast \mathcal{F}_{in} \right ) \right ),
    \label{Eq:mult3}
\end{equation}
where $\bigoplus$ represents the tensor addition operation. Shortcut connection $T_5$ improves network performance and eases training through residual learning. 



\subsection{Dice Similarity Coefficient Loss}
A standard approach to deal with the inherent class imbalance nature of medical images is the reweighting approach based on the frequency of pixels \cite{Eigen2015}. This tackles the imbalance problem at a global level as opposed to a local image-by-image level. Hence, we employ a local technique that maximises the pixel-level Dice similarity coefficient (DSC) between the predictions and the reference on an image-by-image basis. Accordingly, we formulate the loss as
\begin{equation}
\mathcal{L}_\text{Dice} = \sum_{{I} \in \mathcal{D}}(1-\text{DSC})^2,
\label{eq:01}
\end{equation}
where ${I}$ is an image in the training dataset $\mathcal{D}$ under consideration. The DSC between the prediction $\hat{y}_p$ and the gold-standard reference $g_p$ for the $p^\text{th}$ pixel can be given as
\begin{equation}
\text{DSC} = \frac{2\sum_{p\in {I}} \hat{y}_p g_p}{\sum_{p\in {I}} {\hat{y}_p}^2 +\sum_{p\in {I}} g_p^2},
\end{equation}
where $\hat{y}$ and $g$ are binary maps with foreground and background pixels denoted by 1 and 0, respectively. 
Differentiating Eq.~(\ref{eq:01}) yields the gradient as follows:
\begin{equation}
\nabla_{\hat{y}_p}\mathcal{L_\text{Dice}} = 2(1-\text{DSC})\frac{\partial \text{DSC}}{\partial \hat{y}_p}.
\end{equation}
The partial derivative of DSC with respect to the $q^\text{th}$ pixel of the prediction can be calculated as:

\begin{equation}
\frac{\partial\text{DSC}}{\partial \hat{y}_q} =
{\frac{2{{g_q}\left( {\displaystyle\sum_{p\in {I}} {\hat{y}_p^2\ + } \sum_{p\in {I}} {g_p^2} } \right) - 4{\hat{y}_q}\left( {\displaystyle\sum_{p\in {I}} {{\hat{y}_p}{g_p}} } \right)}}{{{{\left( {\displaystyle\sum_{p\in {I}} {\hat{y}_p^2\ + } \sum_{p\in {I}} {g_p^2} } \right)}^2}}}}.
\end{equation}

\section{Experimental Setup}
\label{experimentalResults}
The proposed network was evaluated on a variety of datasets from different medical imaging modalities and domains to examine its broad applicability and performance compared to the state-of-the-art. This section provides an overview of the datasets, the specifics of the network implementation, the performance metrics used, and the comparison methods.

\subsection{Medical Image Datasets}
Seven publicly available datasets from five medical image segmentation applications were used to evaluate the performance of our network \textcolor{black}{(Table \ref{tab:datasets})}. The five applications were segmentation of retinal vessels in fundus images (fundus imaging), segmentation of skin lesions in clinical photographs (optical imaging), segmentation of polyps in intestinal tract colon images (endoscopy imaging), lung segmentation in chest radiographs (X-ray imaging) and segmentation of cell nuclei in histopathological images (microscopic imaging).

\begin{table*}[!t]
    \centering\color{black}
    \begin{tabular}{lllll}
        \hline
        \textbf{Application} & \textbf{Dataset} & \textbf{Image Resolution} & \textbf{Total} & \textbf{Training/Test Split} \\
        \hline
        Retinal Vessels & DRIVE & 584$\times$565 & 40 &Train: 20, Test: 20 \\
        Retinal Vessels & CHASE & 999$\times$960 & 28 &Train: 70\%, Test: 30\% \\
        Skin Lesions & ISIC 2016 & 679$\times$453--6,748$\times$4,499 & 1,279 & Train: 900, Test: 379 \\
        Skin Lesions & ISIC 2017 & 679$\times$453--6,748$\times$4,499 & 2,750 & Train: 2,000, Test: 600 \\
        Polyps & CVC-ClinicDB & 384$\times$288 & 612 & Train: 80\%, Test: 10\%\\
        Lungs & Montgomery County (MC) & 4,892$\times$4,020--4,020$\times$4,892 & 138 & Train: 100, Test: 38\\
        Cell Nuclei & MoNuSeg & 1,000$\times$1,000 pixels & 44 &Train: 30, Test: 14 \\
        \hline
    \end{tabular}
    \caption{Datasets used in the study.}
    \label{tab:datasets}
\end{table*}

For segmentation of retinal vessels, two publicly available retinal image datasets (with a high frequency of containing tiny objects) were used, namely DRIVE and CHASE \cite{Staal2004}. The diabetic retinopathy screening programme that served as the basis for the DRIVE dataset was carried out in the United Kingdom and included patients of a wide age spectrum who were diabetic. It includes 20 images for training and 20 for testing, all sized at 584$\times$565 pixels. Only seven out of the 40 images show minor indications of early diabetic retinopathy. Gold-standard reference manual vessel segmentations by medical experts are available for both training and testing images. The CHASE dataset consists of 28 colour images taken from 14 English schoolchildren. Each image has a resolution of 999$\times$960 pixels and a 30-degree FOV centered on the optic disc. Two distinct manual segmentation maps are provided as a reference, the first expert segmentation used as a reference for this study. The original dataset does not have segregation of images into train and test sets. In our study, the first 70\% images were used for training, and the rest of the images were used for performance evaluation.

For skin lesion segmentation, two datasets were used, namely the ISIC 2017 \cite{8363547} and ISIC 2016 \cite{codella2019skin} datasets. The ISIC 2017 dataset contains 2,750 image sizes ranging from $679\times 453$ to $6,748\times 4,499$ pixels, divided into a training set of 2,000 images, a validation set of 150 images, and a test set of 600 images. Similarly, the ISIC 2016 dataset comprises 1,279 images with sizes ranging from $679\times 453$ to $6,748\times4,499$ pixels. The dataset is split into a training set consisting of 900 images and a test set comprising 379 images.

The CVC-ClinicDB dataset \cite{Bernal2015} was employed for polyp segmentation. This dataset consists of 612 images with a resolution of $384\times288$ pixels, each with a corresponding labelled map. Following the method used in earlier studies \cite{Fan2020,8959021}, we randomly chose 80\% of the images to be used for training, 10\% for testing, and the last 10\% for validation.

The Montgomery County (MC) dataset \cite{Jaeger2014} was used for lung segmentation. This dataset consists of 138 frontal chest X-ray images specifically intended for lung segmentation. These images are either 4,020$\times$4,892 or 4,892$\times$4,020 pixels in size and are presented in 12-bit greyscale. The dataset includes 58 cases of tuberculosis and 80 cases of healthy individuals, highlighting a wide range of anomalies.

The MoNuSeg dataset \cite{liang2015recurrent} comprises 30 training images and 14 test images, each with a resolution of $1,000\times1,000$ pixels. The images are sampled from a diverse array of whole slide images from various organs, thus offering a broad range of examples on which the machine learning model can be trained. To supplement and enrich the dataset, a patch extraction strategy was implemented. In this process, each image is divided to create four unique patches, each of which has dimensions of 512$\times$512 pixels. Patches are generated by extracting content from the four corners of each individual image. This technique significantly increases the size of the original dataset, multiplying it by a factor of four. As a result, instead of the initial set of 44 images, we obtained a considerably expanded dataset containing 176 patches, of which 120 were used for training and 56 for testing. 

 
\subsection{Implementation Details}
Our experiments were carried out on an Intel(R) Core(TM) i9-9900X CPU @ 3.50GHz, equipped with 64 GB of RAM and an NVIDIA Quadro RTX 5000 GPU. We employed the Adam optimisation algorithm with a learning rate of $1e^{-3}$. The model's objective function was the pixel-wise Dice loss function. To address computational constraints, we employed a batch size of 8. Early stopping was integrated to automatically determine the maximum number of iterations, ensuring efficient training without overfitting. Our experiments converged in a maximum of 15 epochs. Gradient clipping was applied using a gradient threshold of 3 and the global $l_2$-norm \cite{Razvan2013jmlr}. This strategy enabled effective experimentation even within the limitations of the available computational resources.

\subsection{Performance Measures}
To assess segmentation performance, we employed several evaluation measures \cite{ Yeghiazaryan-2018}. Not all measures were used for the five applications considered in our experiments. For each application, we used only those measures typically reported in the literature for that application. Given the predicted image, denoted by $P$, and the corresponding gold-standard reference image, denoted by $R$, each consisting of $N$ pixels $i=1\dots N$ with values either 1 (positive $=$ foreground) or 0 (negative $=$ background), we initially calculated the true-positives ($T_P$), true-negative ($T_N$), false-positive ($F_P$), and false-negative ($F_N$) pixel counts:

\begin{equation}\label{eq:tp}
	T_P = \sum^N_{i=1} P_i R_i,
\end{equation}

\begin{equation}\label{eq:tn}
	T_N = \sum^N_{i=1} (1-P_i)(1-R_i),
\end{equation}

\begin{equation}\label{eq:fp}
	F_P = \sum^N_{i=1} P_i (1-R_i),
\end{equation}

\begin{equation}\label{eq:fn}
	F_N = \sum^N_{i=1} (1-P_i) R_i.
\end{equation}

From these calculations, we obtain the sensitivity ($Se$), also known as the Recall:
\begin{equation}\label{eq:se}
	Se = \frac{T_P}{T_P+F_N},
\end{equation}
the specificity ($Sp$):
\begin{equation}\label{eq:sp}
	Sp = \frac{T_N}{T_N+F_P},
\end{equation}
the accuracy ($A$):
\begin{equation}\label{eq:acc}
	A = \frac{T_P+T_N}{T_P+T_N+F_P+F_N},
\end{equation}
the $F_1$ score:
\begin{equation}\label{eq:f1}
	F_1 = \frac{2|P\cap R|}{|P|+|R|} = \frac{2T_P}{2T_P+F_P+F_N},
\end{equation}
and the Jaccard ($J$) coefficient:
\begin{equation}\label{eq:jaccard}
	J = \frac{|P\cap R|}{|P\cup R|} = \frac{T_P}{T_P+F_P+F_N},
\end{equation}
where all values are within the range $[0, 1]$, with 1 indicating optimal performance and 0 representing the worst possible outcome. The area under the curve (\textit{AUC}) of the $Se$-$Sp$ plot is also sometimes used as a performance measure.

An alternative to the $F_1$ score, used in some of the experiments, is the weighted $F$ score:
\begin{equation}\label{F_beta}
F_{\beta}^{w}=(1+\beta^{2})\frac{{\rm Precision}^{w}\cdot{\rm Recall}^{w}}{\beta^{2}\, {\rm Precision}^{w}+{\rm Recall}^{w}},
\end{equation}
where ${\rm Precision} = T_P/(T_P + F_P)$, ${\rm Recall}=Se$, $\omega$ assigns weights to different errors based on neighbourhood information, and $\beta$ is a parameter to balance precision and recall. The weighted precision measures the exactness, while the weighted recall measures the completeness \cite{6909433}. We gave equal importance to weighted precision and recall in our evaluation, which means $\beta = 1$. Additionally, to assess the deviation between the predicted and the reference segmentation at the pixel level, the mean absolute error (\textit{MAE}) was employed:
\begin{equation}
\label{MAE}
\textit{MAE} = \frac{1}{N}\sum_{i=1}^{N} \left|P_i - R_i\right|.
 \end{equation}
For global-level pixel similarity, the enhanced alignment measure $E_{\phi}^{\max}$ was used \cite{Fan2018EM}, given as:
\begin{equation}\label{Emax}
E_{\phi}^{\max} = \frac{1}{K L} \sum_{p=1}^{K} \sum_{q=1}^{L} \phi(p,q),
 \end{equation}
where $\phi = f(\Lambda )$, with $f$ being a quadratic function of the alignment matrix $\Lambda$ \cite{Fan2018EM}. Finally, to quantify structural similarities (\textit{SSIM}) between the predicted and the reference segmentations, the measure $S_\alpha$ \cite{Cheng2021} was used:
\begin{equation}\label{Salfa}
S_\alpha  =\alpha\,S_O + (1-\alpha)\,S_R,
 \end{equation}
where $S_O = \mu\,O_F + (1-\mu)\,O_B$ and $S_R = \sum_{n} w_n\,\textit{SSIM}(n)$. Here, $S_O$ is the object-level similarity, $\mu$ is the ratio of the foreground in the reference segmentation to the image area, and $O_F$ and $O_B$ are the object-level similarities between the predicted and reference segmentation in terms of foreground and background, respectively. The similarity at the region level, $S_R$, is calculated as the weighted foreground ($w_n$) region similarity $\textit{SSIM}$, where the latter is the product of three comparison factors including luminance, contrast, and structure \cite{Cheng2021}. In our evaluation, we used $\alpha=0.5$ to calculate $S_\alpha$. For more details on the measures, we refer to the cited works.

\section{Results and Discussion}
\label{Results}
An ablation study was first performed to evaluate the effectiveness of integrating DMR blocks in the skip connections of the proposed method. This study involved comparing the performance of our method with and without DMR. For this pivotal ablation study, we strategically selected the ISIC 2016 and ISIC 2017 datasets, which focus specifically on dermatology and skin lesion analysis. This decision was influenced by two factors. First, skin image analysis is a field with specific challenges, such as the need for accurate detection and diagnosis of various skin lesions. As skin maladies manifest in a variety of ways, these obstacles make skin-related datasets ideal for evaluating the efficacy of our approach. Additionally, because skin image analysis is very complicated, adding DMR blocks has many advantages, such as better information flow, better gradient propagation, a larger receptive field, and the effective use of fine-grained features. Furthermore, the deep architecture of our network, which is characterised by a large number of branches and separable convolution in-depth, enables it to capture intricate high-frequency details and boundary structures, qualities that are essential for accurate segmentation of skin lesions. As we investigate our findings more thoroughly (see Table \ref{Ablation}), it becomes clear that the incorporation of DMR blocks substantially improves the overall performance of our proposed technique.


\begin{table}[!htbp]
  \centering
    \resizebox{0.7\textwidth}{!}{%
    \begin{tabular}{clrrrrr}
   \toprule
    \multicolumn{1}{l}{\textbf{Dataset}} & \textbf{Method} & \multicolumn{1}{c}{\textbf{Se}} & \multicolumn{1}{c}{\textbf{Sp}} & \multicolumn{1}{c}{\textbf{A}} & \multicolumn{1}{c}{$\textbf{F}_\mathbf{1}$} & \multicolumn{1}{c}{\textbf{J}} \\
    \midrule
    \multirow{2}[2]{*}{ISIC 2016} & Without DMR & 0.9391 & \textbf{0.9732} & 0.9635 & 0.9410 & 0.8879 \\
          & With DMR & \textbf{0.9415} & 0.9688 & \textbf{0.9738} & \textbf{0.9451} & \textbf{0.9018} \\
    \midrule
    \multirow{2}[2]{*}{ISIC 2017} & Without DMR & 0.8730 & 0.9311 & 0.9433 & 0.8741 & 0.7955 \\
          & With DMR & \textbf{0.9049} & \textbf{0.9561} & \textbf{0.9504} & \textbf{0.9034} & \textbf{0.8356} \\
    \bottomrule
    \end{tabular}}%
  \caption{Effect of the proposed DMR block on the performance of the proposed network. Best results are in bold.}
  \label{Ablation}%
\end{table}%


\label{sct:results}
\begin{table*}[!htbp]
  \centering
   \resizebox{1\textwidth}{!}{%
    \begin{tabular}{lcccccc}
    \toprule
    \textbf{Method} & \textbf{Se} & \textbf{Sp} & \textbf{A} & \textbf{AUC} & $\textbf{F}_\mathbf{1}$ & \textbf{Params (M)} \\
    \midrule
    SegNet \cite{khan2020residual} & 0.7949 & 0.9738 & 0.9579 & 0.9720 & 0.8182 & 28.40 \\
    Three-Stage FCN \cite{8476171} & 0.7631 & 0.9820 & 0.9538 & 0.9750 & - & 20.40 \\
    Image BTS-DSN \cite{GUO2019105} & 0.7800  & 0.9806 & 0.9551 & 0.9796 & 0.8208 & 7.80 \\
    VessNet \cite{Arsalan2019}  &   0.8022  & 0.9810  & 0.9655  &0.9820  & - & 9\\
    DRIU \cite{10.1007/978-3-319-46723-8_17} & 0.7855 & 0.9799 & 0.9552 & 0.9793 & 0.8220 & 7.80 \\
    MobileNet-V3 \cite{Howard_2019_ICCV} (Lightweight) & 0.8250 & 0.9771 & 0.9371 & 0.9376 & 0.6575   & 2.50 \\
    ERFNet \cite{Romera_2018} (Lightweight) & - & - & 0.9598 & 0.9598 & 0.7652   & 2.06 \\
    M2U-Net \cite{LaibacherWJ19_CVPRW} (Lightweight) & - & - & 0.9630 & 0.9714 & 0.8091   & 0.55 \\
    Vessel-Net \cite{Wu2019} (Lightweight) & 0.8038 & 0.9802 & 0.9578 & 0.9821 & -   & 1.70 \\
    MS-NFN \cite{Wu2018} (Lightweight) & 0.7844 & 0.9819 & 0.9567 & 0.9807 & -  & 0.40 \\
    FCN \cite{OLIVEIRA2018229} (Lightweight) & 0.8039 & 0.9804 & 0.9576 & 0.9821 & - & 0.20 \\
    T-Net \cite{khan2022t} (Lightweight) & 0.8262 &  \textbf{0.9862} & {0.9697} & \textbf{0.9867} & 0.8269   & \textbf{0.03} \\
    Proposed ESDMR-Net (Lightweight) & \textbf{0.8320} & 0.9832 & \textbf{0.9699} & 0.9853 &\textbf{0.8287} & 0.70 \\
    \bottomrule
    \end{tabular}}
    \caption{Results on the DRIVE dataset. Best results are in bold. Unknown results are indicated by a dash.}
    \label{tab:DRIVE}
\end{table*}

   

\begin{table*}[!htbp]
  \centering
  \resizebox{0.8\textwidth}{!}{%
    \begin{tabular}{lcccccc}
    \toprule
    \textbf{Method} & \textbf{Se} & \textbf{Sp} & \textbf{A} & \textbf{AUC} & $\textbf{F}_\mathbf{1}$ &\textbf{Params (M)} \\
    \midrule
    SegNet \cite{khan2020residual} & 0.8190 & 0.9735 & 0.9638 & 0.9780 & 0.7981 & 28.40\\
    Three-stage FCN \cite{8476171} & 0.7641 & 0.9806 & 0.9607 & 0.9776 & - & 20.40  \\
    Image BTS-DSN \cite{GUO2019105} & 0.7900 & 0.9800 & 0.9630 & 0.9840 & 0.7980 & -\\
    DEU-Net \cite{Wang2019} & 0.8074 & 0.9821 & 0.9661 & 0.9812 & 0.8037 & - \\
    MobileNet-V3 (Lightweight)  \cite{Howard_2019_ICCV} & 0.8121 & 0.9601 & 0.9571 & 0.9673 & 0.6837& 2.50\\
     M2U-Net (Lightweight) \cite{LaibacherWJ19_CVPRW}  & - & - & 0.9703 & 0.9666 & 0.8006& 0.55\\
    Vessel-Net (Lightweight) \cite{Wu2019}& 0.8132 & 0.9814 & 0.966 1& 0.9860 & - & 1.70 \\
    MS-NFN (Lightweight) \cite{Wu2018} & 0.7538 & 0.9847 & 0.9637 & 0.9825 & - & - \\
    DPN (Lightweight) \cite{guo2020dpn} & 0.8360 & 0.9701 & 0.9580 & 0.9781 & 0.7790& 2.40 \\
    T-Net (Lightweight)  \cite{khan2022t}  & 0.8323 & 0.9844 & 0.9739 & \textbf{0.9889} & 0.8143& \textbf{0.03} \\
    Proposed ESDMR-Net (Lightweight) & \textbf{0.8380} & \textbf{0.9850} & \textbf{0.9749} & {0.9881} & \textbf{0.8211}& 0.70 \\
    \bottomrule
    \end{tabular}}
    \caption{Results on the CHASE dataset. Best results are in bold. Unknown results are indicated by a dash.}
    \label{tab:CHASEDB1}
\end{table*}

\begin{figure}[!htbp]
    \centering
    \resizebox{1\textwidth}{!}{%
      \centering
  \includegraphics[width=0.25\textwidth]{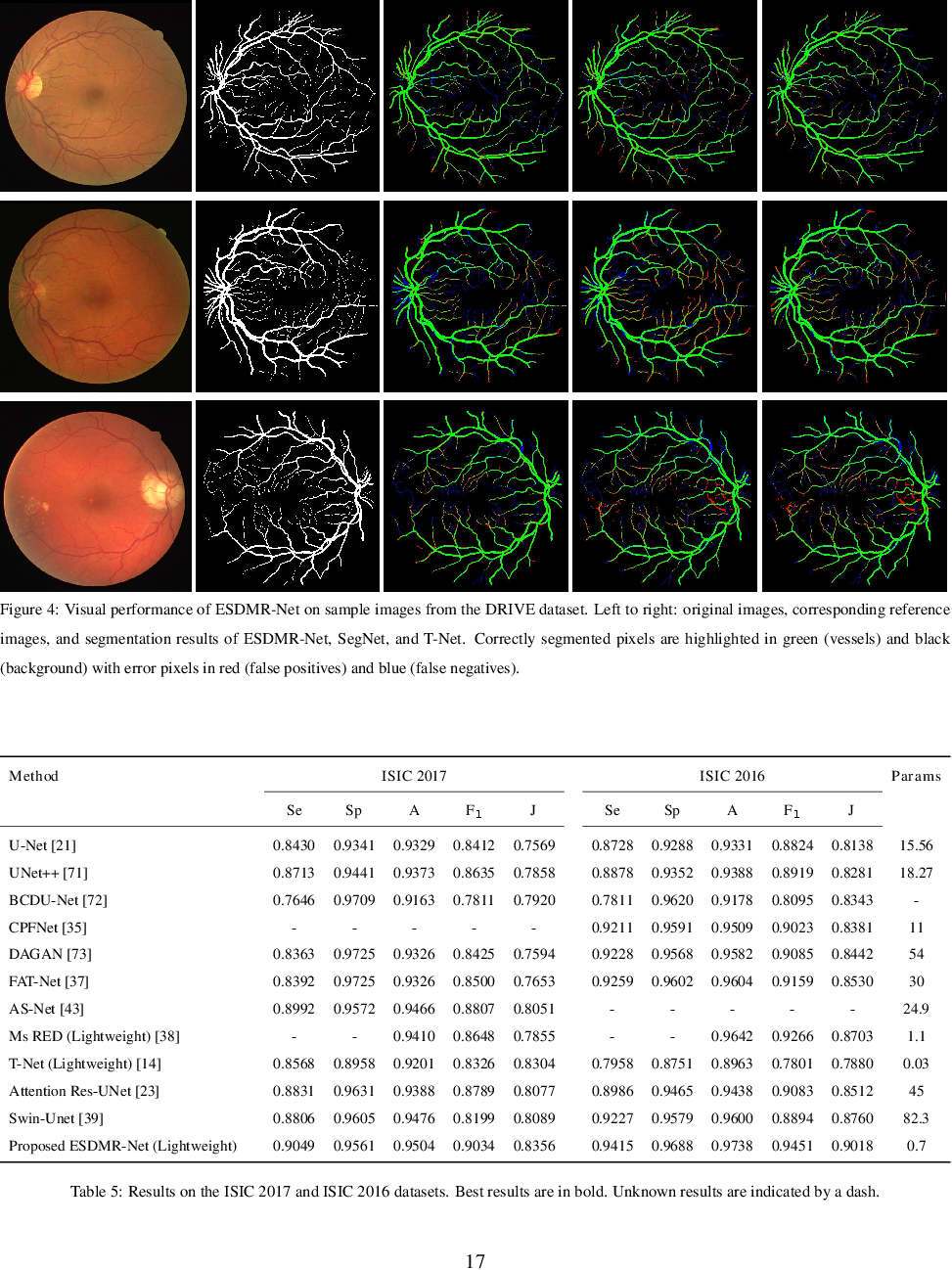} }
    \caption{Visual performance of ESDMR-Net on sample images from the DRIVE dataset. \color{black} Left to right: original images, corresponding reference images, and segmentation results of ESDMR-Net, SegNet, and T-Net. \color{black} Correctly segmented pixels are highlighted in green (vessels) and black (background) with error pixels in red (false positives) and blue (false negatives).}
    \label{fig:DRIVE}
    \vspace{-0.3cm}
\end{figure}

\begin{figure}[!htbp]
    \centering
    \centering
    \resizebox{1\textwidth}{!}{%
      \centering
  \includegraphics[width=0.25\textwidth]{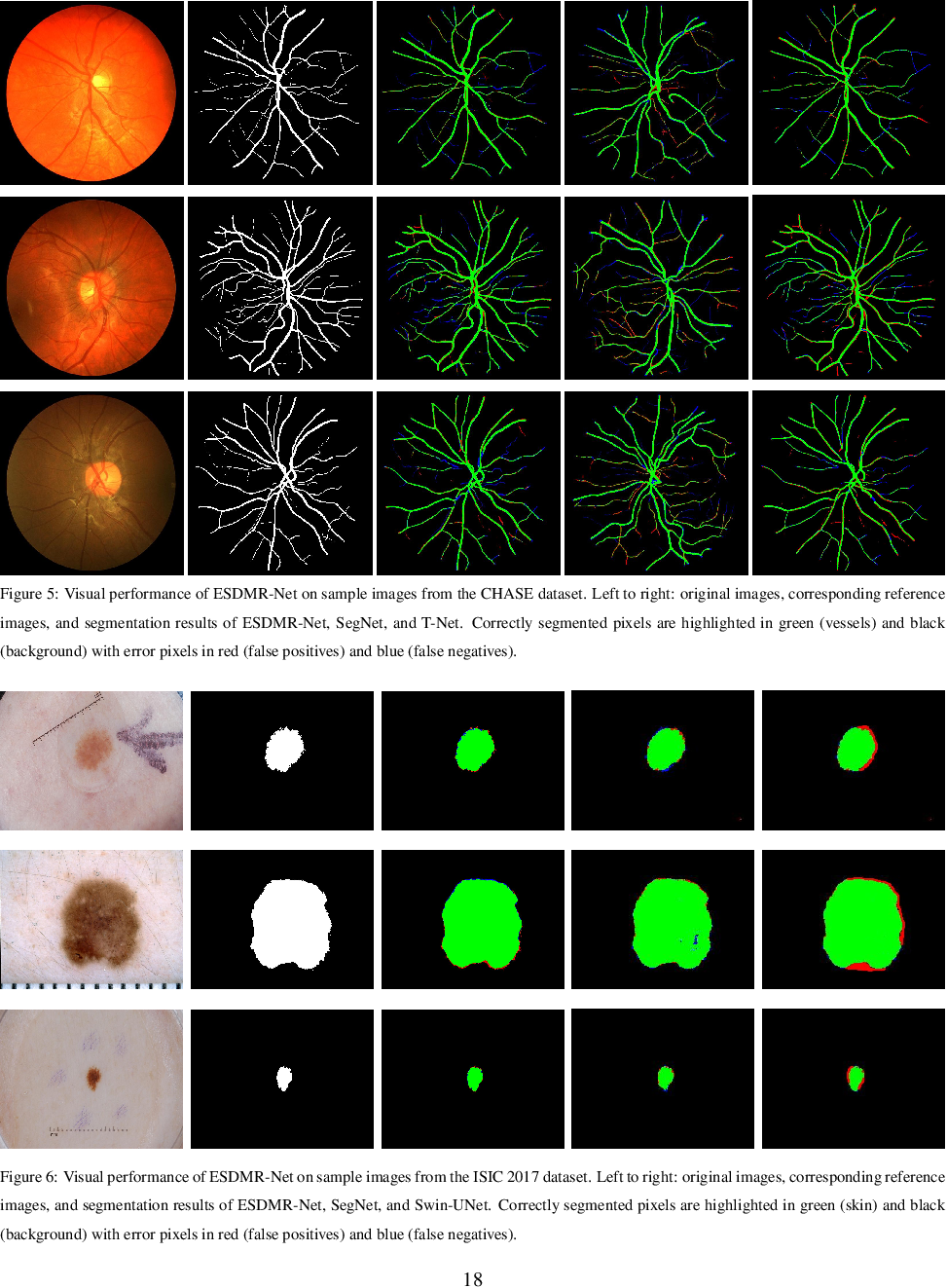} }
    \caption{Visual performance of ESDMR-Net on sample images from the CHASE dataset. \color{black} Left to right: original images, corresponding reference images, and segmentation results of ESDMR-Net, SegNet, and T-Net. \color{black} Correctly segmented pixels are highlighted in green (vessels) and black (background) with error pixels in red (false positives) and blue (false negatives).}
    \label{fig:CHASE}
    \vspace{-0.3cm}
\end{figure}

For the segmentation of the retinal vessels, we performed a comprehensive performance comparison of our ESDMR-Net with 12 other networks previously evaluated in the literature on the DRIVE dataset. Furthermore, we compared our network with 11 other networks on the CHASE dataset. This allowed us to assess the effectiveness of our ESDMR-Net in relation to other state-of-the-art methods across multiple datasets. In the literature on this segmentation task, five performance measures are typically reported, namely $Se$, $Sp$, $A$, \textit{AUC}, and $F_1$, and for comparison networks, we copied their scores for these measures from the articles cited. The results for DRIVE (Table \ref{tab:DRIVE}), and CHASE (Table \ref{tab:CHASEDB1}) clearly show the superiority of ESDMR-Net for the task, despite the fact that our network has a smaller number of parameters than almost all the alternatives. In particular, the proposed network outperformed other lightweight networks such as MS-NFN \cite{Wu2018}, FCN \cite{OLIVEIRA2018229}, Vessel-Net \cite{Wu2019}, and T-Net \cite{khan2022t}. Visual inspection of the ESDMR-Net segmentation results on DRIVE (Fig.~\ref{fig:DRIVE}), and CHASE (Fig.~\ref{fig:CHASE}) revealed, not surprisingly, that most segmentation errors occur in tiny peripheral vessels.

\begin{table*}[!htbp]
  \centering
  \caption{Results on the ISIC 2017 and ISIC 2016 datasets. Best results are in bold. Unknown results are indicated by a dash.}
    \resizebox{1\textwidth}{!}{%
    \begin{tabular}{lccccccccccccc}
    \toprule
\textbf{Method} &        & \multicolumn{5}{c}{\textbf{ISIC 2017}} &       & \multicolumn{5}{c}{\textbf{ISIC 2016}} & \textbf{Params} \\
\cmidrule{3-7}\cmidrule{9-13}      &       & \textbf{Se} & \textbf{Sp} & \textbf{A} & $\textbf{F}_\mathbf{1}$ & \textbf{J} &       & \textbf{Se} & \textbf{Sp} & \textbf{A} & $\textbf{F}_\mathbf{1}$ & \textbf{J} & \\
\cmidrule{1-7}\cmidrule{9-13}  U-Net \cite{10.1007/978-3-319-24574-4_28}  &       & 0.8430 & 0.9341 & 0.9329 & 0.8412 & 0.7569 &       & 0.8728 & 0.9288 & 0.9331 & 0.8824 & 0.8138 & {15.56} \\ 
    UNet++ \cite{zhou2018}   &    & 0.8713 & 0.9441 & 0.9373 & 0.8635 & 0.7858 &       & 0.8878 & 0.9352 & 0.9388 & 0.8919 & 0.8281 & {18.27} \\
    BCDU-Net \cite{azad2019bi} &       & 0.7646 & 0.9709 & 0.9163 & 0.7811 & 0.7920 &       & 0.7811 & 0.9620 & 0.9178 & 0.8095 & 0.8343 & {-} \\
    CPFNet \cite{9049412} &        & -     & -     & -     & -     & -     &       & 0.9211 & 0.9591 & 0.9509 & 0.9023 & 0.8381 & {11} \\
    DAGAN \cite{LEI2020101716}  &        & 0.8363 & \textbf{0.9725} & 0.9326 & 0.8425 & 0.7594 &       & 0.9228 & 0.9568 & 0.9582 & 0.9085 & 0.8442 & {54} \\
    FAT-Net \cite{WU2022102327} &        & 0.8392 & \textbf{0.9725} & 0.9326 & 0.8500 & 0.7653 &       & 0.9259 & 0.9602 & 0.9604 & 0.9159 & 0.8530 & {30} \\
    AS-Net \cite{HU2022117112}  &       & 0.8992 & 0.9572 & 0.9466 & 0.8807 & 0.8051 &       & -     & -     & -     & -     & -     & {24.9} \\
     Ms RED (Lightweight) \cite{DAI2022102293} &          & -     & -     & 0.9410 & 0.8648 & 0.7855 &       & -     & -     & 0.9642 & 0.9266 & 0.8703 & {1.1} \\
    T-Net (Lightweight)  \cite{khan2022t}  &        & 0.8568 & 0.8958 & 0.9201 & 0.8326 & 0.8304 &       & 0.7958 & 0.8751 & 0.8963 & 0.7801 & 0.7880 & {0.03} \\
    Attention Res-UNet \cite{maji2022attention} &        & 0.8831 & 0.9631 & 0.9388 & 0.8789 & 0.8077 &       & 0.8986 & 0.9465 & 0.9438 & 0.9083 & 0.8512 & {45} \\
    Swin-Unet \cite{cao2023swin} &       & 0.8806 & 0.9605 & 0.9476 & 0.8199 & 0.8089 &       & 0.9227 & 0.9579 & 0.9600 & 0.8894 & 0.8760 & {82.3} \\
    Proposed ESDMR-Net (Lightweight) &       & \textbf{0.9049} & {0.9561} & \textbf{0.9504} & \textbf{0.9034} & \textbf{0.8356} &       & \textbf{0.9415} & \textbf{0.9688} & \textbf{0.9738} & \textbf{0.9451} & \textbf{0.9018} & \textbf{0.7} \\
    \bottomrule
    \end{tabular}%
    }
  \label{tab:ISIC}%
\end{table*}

\begin{figure}[!htbp]
          \centering
    \resizebox{1\textwidth}{!}{%
      \centering
  \includegraphics[width=1\textwidth]{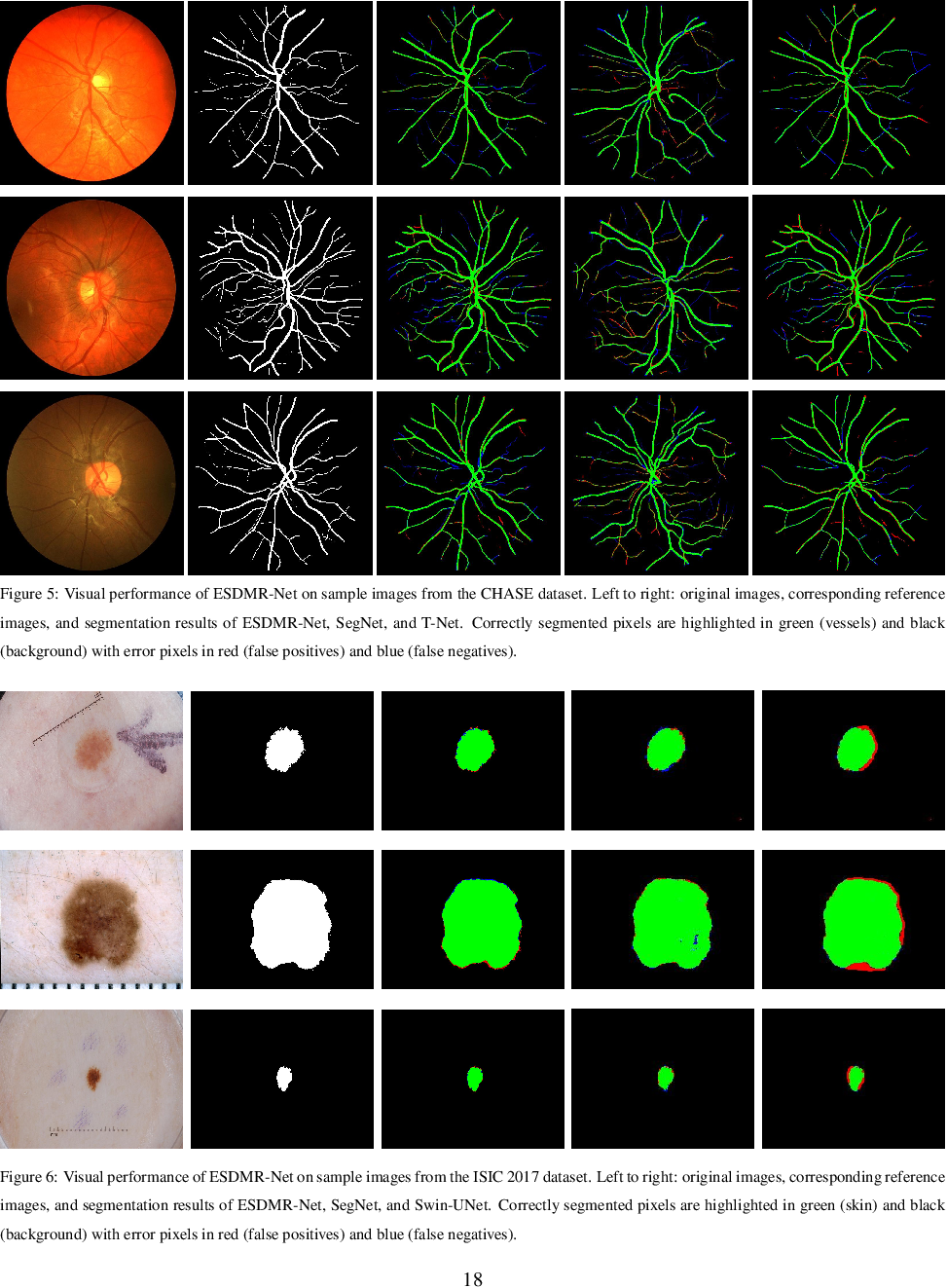} }
    \caption{Visual performance of ESDMR-Net on sample images from the ISIC 2017 dataset. \color{black} Left to right: original images, corresponding reference images, and segmentation results of ESDMR-Net, SegNet, and Swin-UNet. \color{black} Correctly segmented pixels are highlighted in green (skin) and black (background) with error pixels in red (false positives) and blue (false negatives).}
    \label{fig:ISIC2017}
    \vspace{-0.3cm}
\end{figure}

 \begin{figure}[!htbp]
     \centering
    \resizebox{1\textwidth}{!}{%
      \centering
  \includegraphics[width=0.25\textwidth]{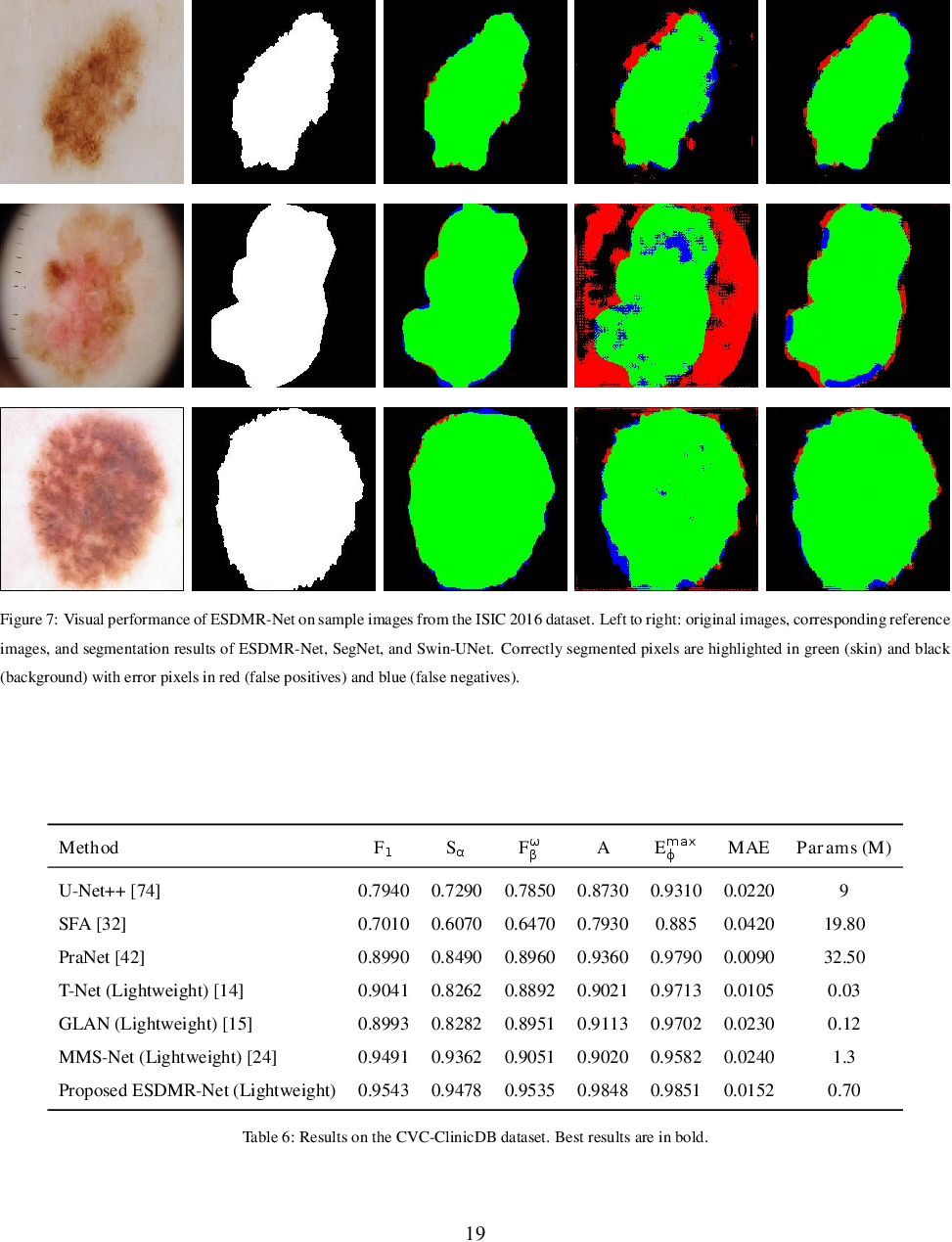} }
    \caption{Visual performance of ESDMR-Net on sample images from the ISIC 2016 dataset. \color{black} Left to right: original images, corresponding reference images, and segmentation results of ESDMR-Net, SegNet, and Swin-UNet. \color{black}  Correctly segmented pixels are highlighted in green (skin) and black (background) with error pixels in red (false positives) and blue (false negatives).}
    \label{fig:SKIN}
    \vspace{-0.3cm}
\end{figure}

\begin{table*}[!htbp]
    \centering
    \resizebox{0.9\textwidth}{!}{%
	\begin{tabular}{lccccccc}
		\toprule
		\textbf{Method} & $\textbf{F}_\mathbf{1}$ & $\boldsymbol{\textbf{S}_\alpha}$ & $\boldsymbol{\textbf{F}_\beta^\omega}$ & \textbf{A} & $\boldsymbol{\textbf{E}_\varphi^{\max}}$
		& \textbf{MAE} & \textbf{Params (M)}\\
		\midrule
		U-Net++ \cite{Zongwei2018} & 0.7940 & 0.7290 & 0.7850 & 0.8730 & 0.9310 & 0.0220& 9 \\
		SFA \cite{Fang2019} & 0.7010 & 0.6070 & 0.6470 & 0.7930 & 0.885 & 0.0420& 19.80 \\
		PraNet \cite{Fan2020} & 0.8990 & 0.8490 & 0.8960 & 0.9360 & 0.9790 & \textbf{0.0090}& 32.50 \\
	    T-Net (Lightweight) \cite{khan2022t} &0.9041 & 0.8262 &0.8892 & 0.9021 & {0.9713} & 0.0105 & \textbf{0.03} \\
            GLAN (Lightweight) \cite{naqvi2023glan} &0.8993 & 0.8282 &0.8951 & 0.9113 & {0.9702} & 0.0230 & 0.12 \\
            MMS-Net (Lightweight) \cite{khan2023simple} &0.9491 & 0.9362 &0.9051 & 0.9020 & {0.9582} & 0.0240 & 1.3 \\
		Proposed ESDMR-Net (Lightweight) & \textbf{0.9543} & \textbf{0.9478} & \textbf{0.9535} & \textbf{0.9848} & \textbf{0.9851} & 0.0152 & 0.70 \\
		\bottomrule
	\end{tabular}}
    \caption{Results on the CVC-ClinicDB dataset. Best results are in bold.}
	\label{tab:CVC}
\end{table*}

\begin{figure}[!htbp]
      \centering
    \resizebox{1\textwidth}{!}{%
      \centering
  \includegraphics[width=0.25\textwidth]{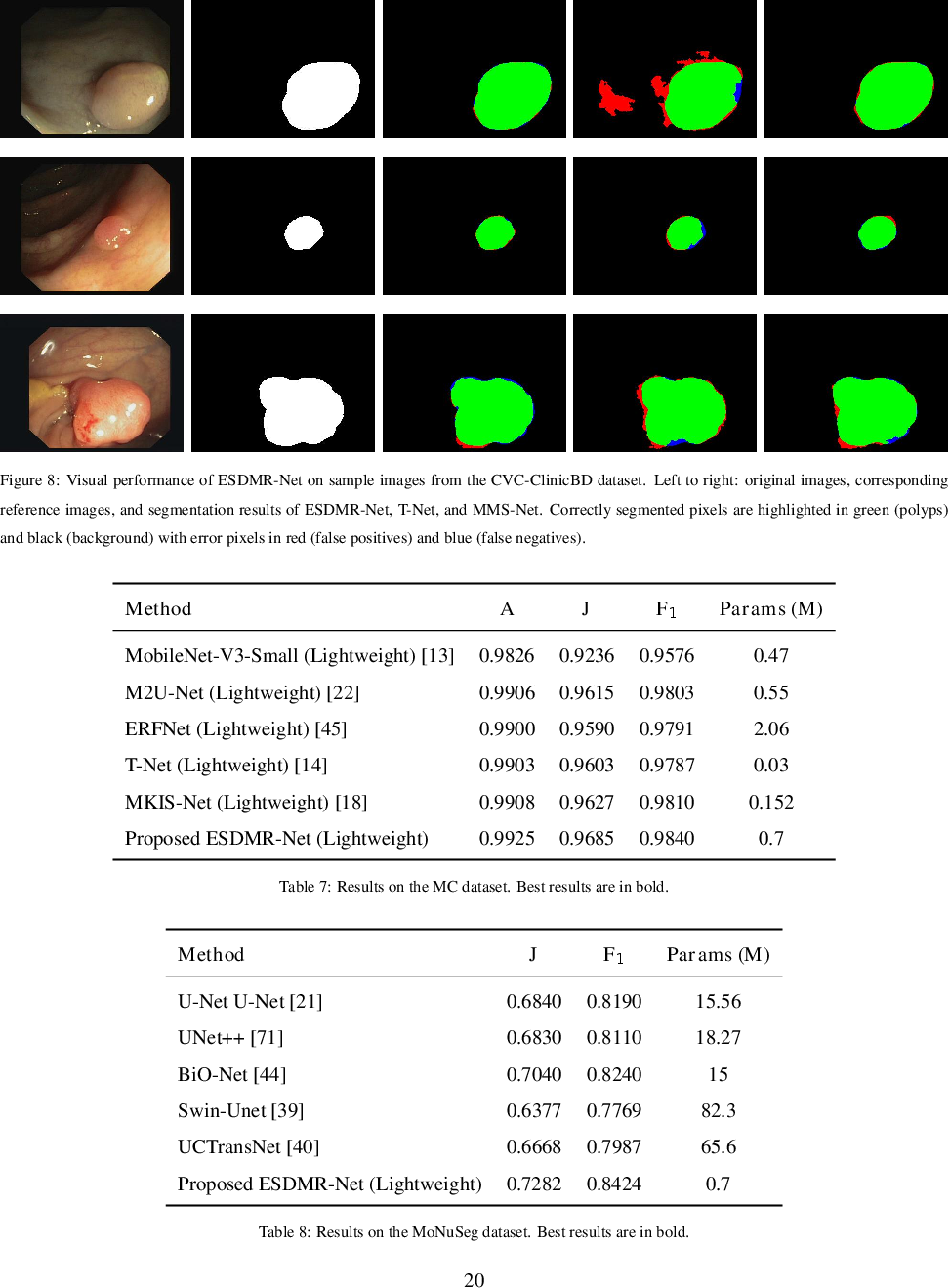} }
    \caption{Visual performance of ESDMR-Net on sample images from the CVC-ClinicBD dataset. \color{black} Left to right: original images, corresponding reference images, and segmentation results of ESDMR-Net, T-Net, and MMS-Net. \color{black} Correctly segmented pixels are highlighted in green (polyps) and black (background) with error pixels in red (false positives) and blue (false negatives).}
    \label{fig:POLYP}
    \vspace{-0.3cm}
\end{figure}

\begin{table*}[!htbp]
\centering
\begin{tabular}{lcccc}
\toprule
\textbf{Method} & \textbf{A} & \textbf{J} & $\textbf{F}_\mathbf{1}$ & \textbf{Params (M)} \\ 
\midrule
MobileNet-V3-Small  (Lightweight)  \cite{Howard_2019_ICCV}& 0.9826 & 0.9236 & 0.9576 & 0.47 \\
M2U-Net (Lightweight) \cite{LaibacherWJ19_CVPRW}   & 0.9906 & 0.9615 & 0.9803 & 0.55 \\
ERFNet (Lightweight) \cite{Romera_2018} & 0.9900 & 0.9590 & 0.9791 & 2.06 \\
T-Net (Lightweight) \cite{khan2022t}  & 0.9903 & 0.9603 & 0.9787 & \textbf{0.03} \\
MKIS-Net (Lightweight) \cite{khan2022mkis}  & 0.9908 & 0.9627 & 0.9810 & {0.152} \\
Proposed ESDMR-Net (Lightweight) & \textbf{0.9925} & \textbf{0.9685} & \textbf{0.9840} & 0.7 \\
\bottomrule
\end{tabular}
\caption{Results on the MC dataset. Best results are in bold.}
\label{tab:MC}
\end{table*}

\begin{figure}[!htbp]
         \centering
    \resizebox{1\textwidth}{!}{%
      \centering
  \includegraphics[width=1\textwidth]{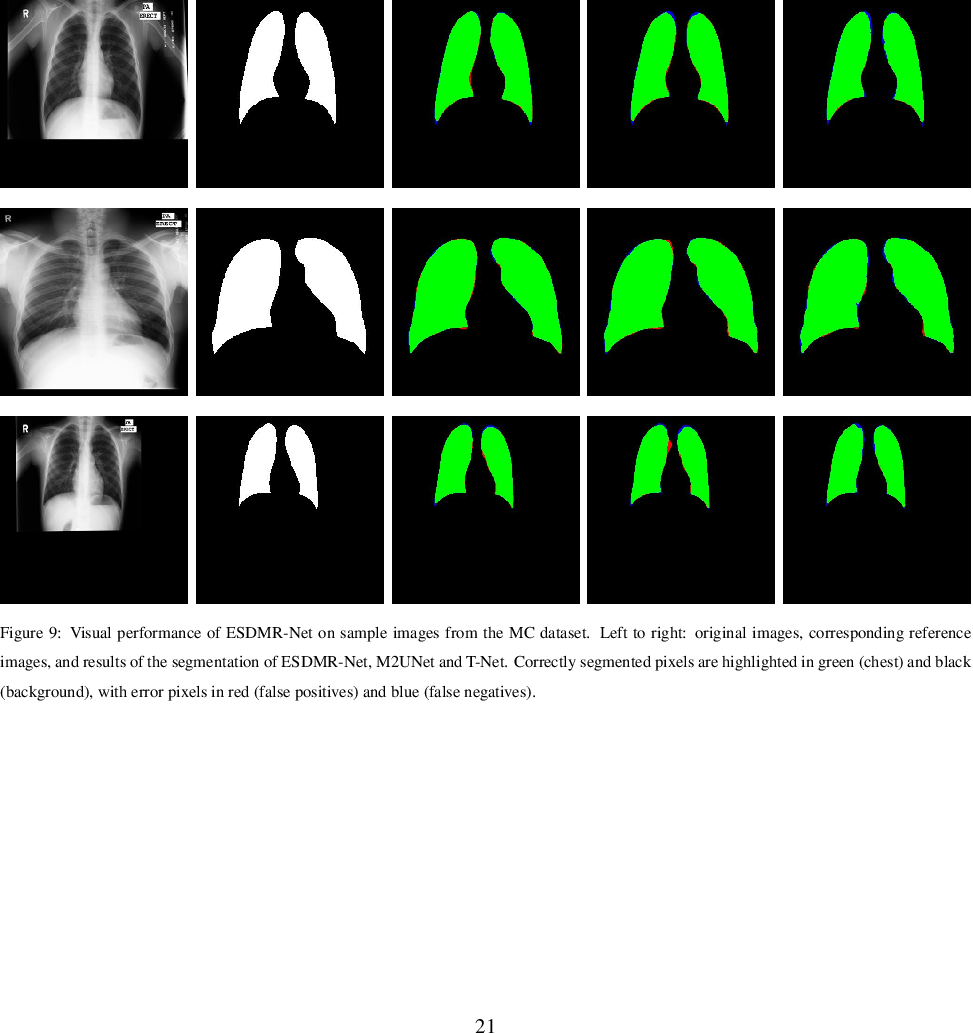} }
    \caption{Visual performance of ESDMR-Net on sample images from the MC dataset. \color{black} Left to right: original images, corresponding reference images, and results of the segmentation of ESDMR-Net, M2UNet and T-Net. \color{black} Correctly segmented pixels are highlighted in green (chest) and black (background), with error pixels in red (false positives) and blue (false negatives).}
    \label{fig:MC}
    \vspace{-0.3cm}
\end{figure}

\begin{table*}[!htbp]
    \centering
    \begin{tabular}{lccc}
    \toprule
    \textbf{Method} & $\textbf{J}$ & $\textbf{F}_\mathbf{1}$ & \textbf{Params (M)} \\
    \midrule
    U-Net U-Net \cite{10.1007/978-3-319-24574-4_28} & 0.6840 & 0.8190 & 15.56 \\
    UNet++ \cite{zhou2018}  & 0.6830 & 0.8110 & 18.27 \\
    BiO-Net \cite{BiO-Net2020} & 0.7040 & 0.8240 & 15 \\
    Swin-Unet \cite{cao2023swin} & 0.6377 & 0.7769 & 82.3 \\
    UCTransNet \cite{UCTransNet2022} & 0.6668 & 0.7987 & 65.6 \\
   Proposed ESDMR-Net (Lightweight)& \textbf{0.7282} & \textbf{0.8424} & \textbf{0.7} \\
    \bottomrule
    \end{tabular}
    \caption{Results on the MoNuSeg dataset. Best results are in bold.}
    \label{tab:MS}
\end{table*}

\begin{figure}[!htbp]
         \centering
    \resizebox{1\textwidth}{!}{%
      \centering
  \includegraphics[width=1\textwidth]{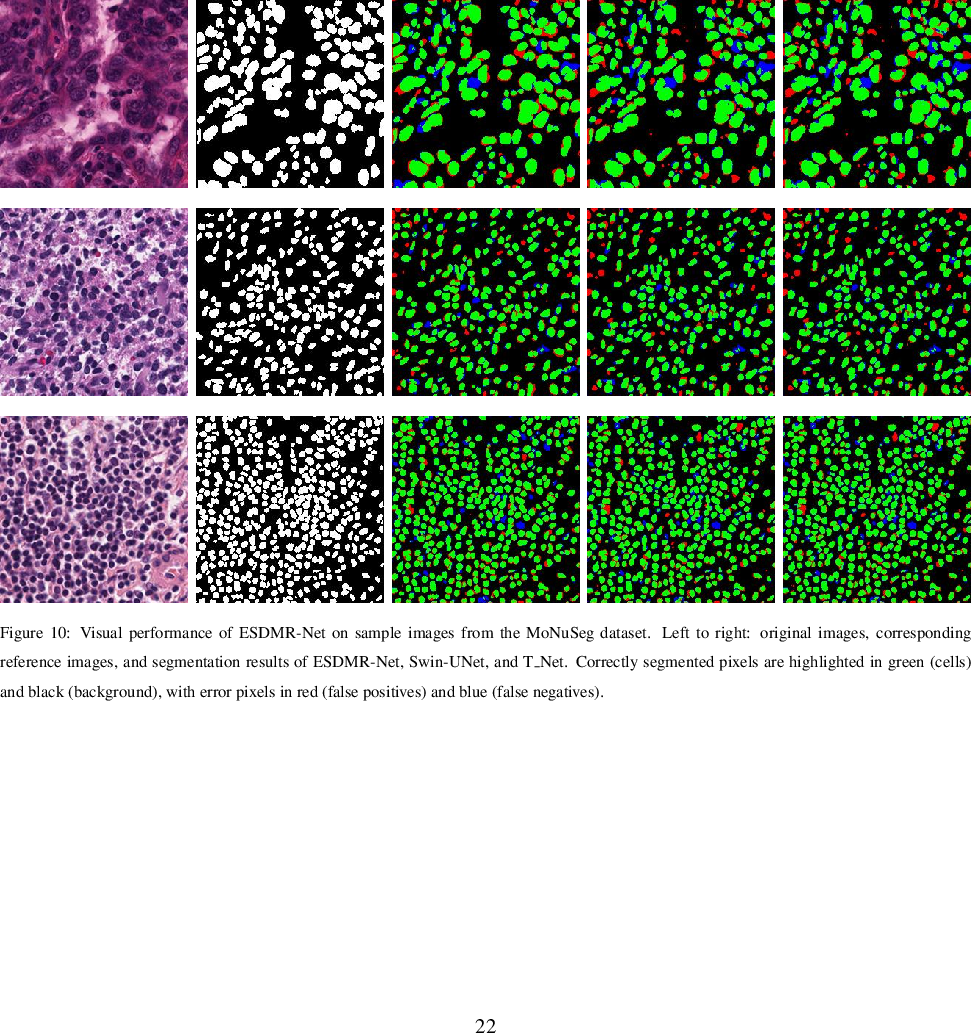} }
    \caption{Visual performance of ESDMR-Net on sample images from the MoNuSeg dataset. Left to right: original images, corresponding reference images, and segmentation results of ESDMR-Net, Swin-UNet, and T\_Net. \color{black} Correctly segmented pixels are highlighted in green (cells) and black (background), with error pixels in red (false positives) and blue (false negatives).}
    \label{fig:MoNuSeg}
    \vspace{-0.3cm}
\end{figure}

\color{black}
For skin lesion segmentation, we compared ESDMR-Net with 10 other networks on the ISIC 2017 and ISIC 2016 datasets. Performances are typically reported in terms of $Se$, $Sp$, $A$, $F_1$, and $J$ in the literature. The results (Table \ref{tab:ISIC}) show that our method outperformed all other networks in terms of $Se$, $A$, $F_1$, and $J$, despite being a lightweight network. From these results and illustrative visual examples from the ISIC 2017 dataset (Fig.~\ref{fig:ISIC2017}) and the ISIC 2016 dataset (Fig.~\ref{fig:SKIN}) we conclude that our method performs well in a wide range of lesion sizes, shapes, colours and textures.

For polyp segmentation, we compared ESDMR-Net with 6 other networks in the CVC-ClinicDB dataset. Here, the performance measures included the mean $F_1$ score, $S_{\alpha}$, $F_{\beta}^{\omega}$, $A$, $E_{\varphi}^{\max}$, and \textit{MAE}. The results (Table \ref{tab:CVC}) again show that ESDMR-Net outperformed other methods in most performance measures. Visual examples (Fig.~\ref{fig:POLYP}) illustrate the high fidelity of the results produced by our network.

For chest X-ray segmentation, we compared ESDMR-Net with four other networks in the MC dataset. Here, the performance measures included the mean $A$, $J$, and $F_1$. The results (Table \ref{tab:MC}) demonstrate that ESDMR-Net exceeded other methods. Furthermore, visual illustrations (Fig.~\ref{fig:MC}) confirm the superior quality of the segmentations by our network.

Finally, for cell image segmentation, we compared ESDMR-Net with five other networks on the MoNuSeg dataset. Here, the performance measures reported included the mean $J$, $F_1$, and the training parameters. The results (Table \ref{tab:MS}) show once again that ESDMR-Net outperforms other methods in terms of most performance measures with visual examples (Fig.~\ref{fig:MoNuSeg}) that confirm the high fidelity of the segmentations produced by our network. Collectively, the results of the considered datasets demonstrate the consistently competitive performance of ESDMR-Net, demonstrating its robustness in a wide variety of medical image segmentation tasks.

\section{\textcolor{black}{Discussion and Conclusions}}

\label{Conclusions}


This paper addresses the challenge of computationally demanding medical image segmentation approaches by proposing the expand-squeeze dual multiscale residual network (ESDMR-Net). The network is designed to be highly efficient on resource-constrained computing hardware, making it well suited for practical applications. ESDMR-Net gets better at segmentation and capturing contextual dependencies by using dual-stream information flow and extracting multiscale features. The ES block in ESDMR-Net plays a crucial role in paying more attention to under-represented classes, contributing to improved segmentation performance. The addition of DMR blocks to skip connections also improves the flow of data across multiple resolutions or scales, allowing the decoder to access features at different levels of abstraction. This integration results in more comprehensive feature representations, further improving the quality of the segmentation. Experiments on five different medical image segmentation applications demonstrate the effectiveness of ESDMR-Net. Although it has significantly fewer trainable parameters compared to other cutting-edge methods, ESDMR-Net outperforms them in terms of segmentation performance. This has significant implications for the development of efficient and accurate segmentation models that can be deployed in resource-constrained settings, ultimately benefiting physicians and improving patient care.

\color{black}
ESDMR-Net distinguishes itself from other medical image segmentation approaches by offering a compelling solution that is notable for its efficiency and lightweight architecture. Conventional methods, such as U-Net variations and those found by neural architecture search techniques, usually need a lot of resources, which means that regular hardware cannot use them. ESDMR-Net performs better than these models due to its low number of parameters and high computational efficiency, making it a good choice for deployment in settings with limited memory and processing power. ESDMR-Net is great at finding the right balance between model complexity and performance. It solves the important problem of medical image segmentation, which other designs might find hard to scale down. Unique additions of the expand-squeeze and dual multiscale residual blocks allow ESDMR-Net to accurately capture complex features and successfully address the class imbalance. This demonstrates its superiority over current approaches in handling high-frequency details and boundary structures in medical pictures.

In terms of limitations and possible future extensions of ESDMR-Net, we note that while our experiments included a variety of medical imaging datasets, further investigation is needed to demonstrate the model's capacity to generalise across other medical imaging applications. Performance could be further improved through task-specific optimisations, while interpretability may be improved by incorporating explainable AI approaches. To prove that ESDMR-Net is a reliable solution in the complicated field of medical image segmentation, it is important to look at how it would be used in real life and compare it to the most advanced models currently available. With continual improvements and extensions, ESDMR-Net is a potential option to improve the ability of medical image analysis in settings with limited resources.

\color{black}

\end{document}